\documentclass{article}

\usepackage{arxiv}

\usepackage[utf8]{inputenc} % allow utf-8 input
\usepackage[T1]{fontenc}    % use 8-bit T1 fonts
\usepackage{hyperref}       % hyperlinks
\usepackage{url}            % simple URL typesetting
\usepackage{booktabs}       % professional-quality tables
\usepackage{amsfonts}       % blackboard math symbols
\usepackage{nicefrac}       % compact symbols for 1/2, etc.
\usepackage{microtype}      % microtypography
\usepackage{lipsum}
\usepackage{graphicx}
\usepackage{threeparttable}
\usepackage{amsmath}
\usepackage{natbib}
\usepackage[dvipsnames]{xcolor}
\usepackage{multicol}
\usepackage{multirow}
\usepackage[all]{hypcap}    % links to image top corner instead of the caption 

\hypersetup{ colorlinks=true, linkcolor={blue}, filecolor=magenta, urlcolor={magenta}, citecolor={blue}}
\newcommand{\mnras}{\textit{Monthly Notices of the Royal Astronomical Society}}
\newcommand{\aap}{\textit{Astronomy and Astrophysics}}
\newcommand{\apjs}{\textit{Astrophysical Journal, Supplement}}
\newcommand{\apj}{\textit{Astrophysical Journal}}
\newcommand{\apjl}{\textit{Astrophysical Journal, Letters}}
\newcommand{\ssr}{\textit{Space Science Reviews}}
\newcommand{\na}{\textit{New Astronomy}}
\newcommand{\memsai}{\textit{Mem. Societa Astronomica Italiana}}
\newcommand{\pasp}{\textit{Publications of the Astronomical Society of the Pacific}}

\title{Multi Frequency Temporal and Spectral variability study of Blazar PKS 1424-418}

\author{
 Jayant Abhir \\
  Department of Physics\\
  IIT Kharagpur\\
   \And
 Jophin Joseph \\
  Department of Physics\\
  IIT Kharagpur\\
  \And
 Sonal R Patel \\
  Department of Physics\\
  University of Mumbai\\
  \And
 Debanjan Bose \\
  Department of Physics\\
  IIT Kharagpur\\
  \texttt{debanjan.tifr@gmail.com}
}

\begin{document}
\maketitle

\begin{abstract}
A study of blazar PKS 1424-418 was carried out using multi waveband data collected by \textit{Fermi}-LAT, \textit{Swift}-XRT, \textit{Swift}-UVOT and \textit{SMARTS} telescopes between MJD 56000 to MJD 56600 (14 Mar 2012 to 4 Nov 2013). Two flaring episodes were identified by analysing the gamma ray light curve. Simultaneous multi waveband Spectral Energy Distributions (SED) were obtained for those two flaring periods. A cross-correlation analysis of IR-Optical and $\gamma$-ray data suggested the origin of these emissions from the same region. We have set a lower limit for the Doppler factor using the highest energy photon observed from this source during the flaring periods, which should be $>$12. The broadband emission mechanism was studied by modelling the SED using leptonic emission mechanism. 
\end{abstract}

% keywords can be removed

\section{Introduction}
Blazars are a class of Active Galactic Nuclei (AGN) with their jets pointed largely in the direction of earth (\cite{1995PASP..107..803U}). At the center of a blazar lies a super massive black hole (SMBH) fueled by the accretion of surrounding matter (\cite{2012agn..book.....B}). The emission in blazars is non-thermal in nature (\cite{2009MNRAS.397..985G}) and they are usually a very strong source of $\gamma$-rays (\cite{1999ApJS..123...79H}). Blazars are mainly divided into two classes B L Lacerate (BLLac) and Flat Spectrum Radio Quasar (FSRQ). BL Lacs do not exhibit any optical emission lines whereas FSRQs do. A blazar consists of the following physical regions: a central SMBH, an accretion disk emitting X-rays, a Broad Line Region (BLR) emitting  highly Doppler broadened photons via atomic transitions, a dust torus emitting mostly in infra-red and a relativistic jet where the high energy $\gamma$-rays are produced (\cite{Celotti2008}). Blazars are highly variable and often exhibit $\gamma$-ray variability ranging from a few days to a few months (\cite{2010ApJ...722..520A}, \cite{2019ApJ...885...12R}).

 Spectral Energy Distributions (SED) obtained with multi waveband data over a wide energy range are powerful tool to understand high energy emission from  blazars (\cite{2010ApJ...716...30A}). By studying the SEDs of blazars one can shed light on the acceleration mechanisms involved in production of high energy cosmic-rays and $\gamma$-rays. These SEDs are characterised by two "humps". According to leptonic models, the first hump (extended over optical-UV-X-ray energies) in the SED is produced due to synchrotron emission by electrons (or positrons) present in the jet. The second hump can be produced in two ways, first, synchrotron photons can get up-scattered by the same electrons (or positrons) producing them (Synchrotron Self Compton (SSC) models, \cite{1985A&A...146..204G}, \cite{1992ApJ...397L...5M})  or these particles can up-scatter external photons, in the BLR, accretion disk or dust torus (External Compton (EC) models, \cite{1992A&A...256L..27D}, \cite{1993ApJ...416..458D}, \cite{1994ApJ...421..153S}). In hadronic models
protons are responsible for high energy emission via proton-proton or proton-photon interactions (\cite{1993A&A...269...67M}, \cite{2000NewA....5..377A}). We have used a leptonic model to fit SEDs of PKS 1424-418 in this work.

PKS 1424-418 is a Flat Spectrum Radio Quasar (FSRQ), as identified in 3FGL catalogue (identifier: J1427.9-4206) (\cite{2017yCat..22320018A}) with a redshift ($z$) of $1.522$ and an estimated SMBH mass of about $4.5 \times 10^9 M_\odot$ (\cite{2004ApJ...602..103F}). It is estimated to be at $11.55$ Gpc distance from earth. Being a FSRQ, the $\gamma$-ray luminosity of PKS 1424-418 dominates its radiative power. The source has previously shown flaring/active states during 2008-2011 (\cite{2014A&A...569A..40B}) and 2012-2013 (\cite{inproceedings}) and has been modelled previously by \cite{Celotti2008} and by \cite{2017ApJ...851...33P} while the blazar was in its average state. PKS 1424-418 has most of its emission from a compact central region and has a low surface brightness jet with wide opening angle, as observed using high resolution Very Long Baseline Interferometry (VLBI) carried under the TANAMI program and its emission is highly polarised (\cite{inproceedings}). The correlation between radio flares and $\gamma$-ray flares for the source has been studied by \cite{2015MmSAI..86...36V}.

In this paper we have studied PKS 1424-418 from 14 Mar 2012 to 4 Nov 2013 using multi-waveband data. In section 2, we have described the data and the analysis techniques. We have presented our results about multi-wavelength variability and flux correlation studies in section 3. Spectral energy distributions and its modelling is described in section 4. We have discussed our results in section 5. Finally a summary is provided in section 6.

\section{Multi-waveband observations and data analysis}
In this section, we describe the multi-waveband data analysis to generate the multi waveband light curve that we have used to identify flaring states of the source. Two flaring periods are identified using the $\gamma$-ray light curve, the first flare (Flare 1) : MJD 56299-56321 and the second flare (Flare 2): MJD 56384-56412. The following sub-sections contain description of the data analysis of $\gamma$-ray data collected by \textit{Fermi} mission, X-ray and UV-optical data collected by \textit{Swift} mission and archival optical band data from the SMARTS telescope.

\subsection{High energy \texorpdfstring{$\gamma$}{gamma}-ray observations of \textit{Fermi}-LAT}
The high energy $\gamma$-ray data collected by \textit{Fermi} satellite's Large Area Telescope (LAT) instrument was used to obtain a light curve and $\gamma$-ray SED for PKS 1424-418 (4FGL J$1427.9-4206$, \cite{Ajello_2020}). \textit{Fermi}-LAT is a pair-conversion space-borne $\gamma$-ray detector launched in June 2008. It has a very large FoV of about 2.4 sr (\cite{2009ApJ...697.1071A}). It is sensitive over a wide energy range, from few tens of MeV to few hundreds of GeV. The energy resolution is about 15$\%$ at 100 MeV range and worse at lower energies due to physical limitations of detection methods. The uncertainty in single photon resolution is $<3.5^\circ$ for energies in the range of 100 MeV and improves to $<0.6^\circ$ for energies greater than 1 GeV (\cite{2009ApJ...697.1071A}). This necessitates a likelihood analysis for identifying potential candidate sources within the region of interest. The analysis was greatly simplified by the use of `\href{https://github.com/fermi-lat/Fermitools-conda}{fermitools}’ (v1.2.1) provided by the \textit{Fermi}-LAT team. The SED was analysed using the user contributed tool `\href{https://enrico.readthedocs.io/en/latest/}{enrico}’ created by \cite{2013ICRC...33.2784S} and the calculated parameters are given in Table \ref{tab:AnalysisResult}.

The \textit{Fermi}-LAT data in the energy range of 0.1–300GeV was collected from MJD 56000 (14 Mar 2012) to MJD 56600 (4 Nov 2013). Events were extracted from the region of interest (ROI) of $15^\circ$ centered around the source position (Right Ascension:  216.985, declination: -42.1054). Photon data files were filtered with ‘\texttt{evclass=128}’ and ‘\texttt{evtype=3}’. In order to avoid the contamination of the data from Earth’s limb, a zenith angle cut of $90^\circ$ was applied. A filter of ‘\texttt{(DATA\_QUAL>0)\&\&(LAT\_CONFIG==1)}’ was applied to select good time intervals. 

The \textit{Fermi}-LAT light curve data was binned at 5 day intervals in the analysis. This choice allowed achieving a test statistic ($TS$) value greater than 9 for the source for all time bins from MJD 56000 to MJD 56600 and $TS$ greater than 900 for the flaring periods (as the flux was higher during flaring period). \texttt{Npred}, which refers to the number of photons in a bin in the likelihood analysis, was $>30$ for all time bins in the light curve. The isotropic and diffuse background were modelled using ‘\texttt{iso\_P8R3\_SOURCE\_V2\_v1.txt}’  and ‘\texttt{gll\_iem\_v07.fits}’ respectively. An unbinned likelihood analysis was performed using \texttt{gtlike} (\cite{1979ApJ...228..939C}, \cite{1996ApJ...461..396M}). Our model contained 229 point sources from the 4FGL catalog (\cite{2020ApJS..247...33A}) which were within $15^\circ$ of PKS 1424-418. The spectral shapes and initial parameters were set using the values published in 4FGL catalog and a total of 11 parameters were kept free in the likelihood analysis. Power Law model was used for the source as given below: 
\begin{equation}
    \frac{dN(E)}{dE} = N_0\times\left(\frac{E}{E_0}\right)^{-\alpha}
\end{equation}
where $E_0$ and $N_0$ are the scale factor and the prefactor respectively provided in the 4FGL catalog and $\alpha$ is the spectral index.The $\gamma$-ray light curve for the source is shown in Figure \ref{fig:LightCurve}.

\subsection{X-ray observations of \textit{Swift}-XRT}
The X-ray data in the energy range 0.3-8 keV, collected by 
the \textit{Swift}-XRT was analysed for the two flaring periods. \href{https://heasarc.gsfc.nasa.gov/docs/software/heasoft/}{HEAsoft} package (v6.26.1) and \href{https://heasarc.gsfc.nasa.gov/xanadu/xspec/}{XSPEC} (v12.10.1f) were used for the analysis of X-ray data.

\textit{Swift}-XRT made 11 observations for this source in the time period from MJD 56300 to 56500. These observations included two flares, from MJD 56299 to 56316 and from MJD 56399 to 56411 respectively. Data was unavailable for rest of the time period of \textit{Fermi}-LAT observations. These flares were found to be almost simultaneous with \textit{Fermi}-LAT flares. First, clean event files were obtained using \href{https://www.swift.ac.uk/analysis/xrt/xrtpipeline.php}{xrtpipeline}. The source spectrum was extracted as Pulse Height Amplitude (PHA) files from the clean event files using XSPEC. The source events were selected from a circular region with a radius of 20 pixels (1 pixel $\sim2.36''$) while the radius was 40 pixels for the background spectrum. The response file was obtained from \href{https://heasarc.gsfc.nasa.gov/docs/heasarc/caldb/caldb_supported_missions.html}{HEASARC calibration database} and the ancillary response files were generated using \href{https://heasarc.gsfc.nasa.gov/ftools/caldb/help/xrtmkarf.html}{xrtmkarf}. The source event file, background event file, response file and ancillary response file were tied together using \href{https://heasarc.gsfc.nasa.gov/lheasoft/ftools/fhelp/grppha.txt}{grppha} and binned together to give a minimum of 20 counts. We did not use pileup corrections since the count rate is well below 0.5 count $s^{-1}$ for all 11 observations. The data was $\chi^{2}$ fitted using XSPEC version 12.10.1f. The data was fitted with a basic power-law ($F(E)= K E^{\Gamma_x}$). For this fit, neutral hydrogen column density was fixed at $N_{H}=7.7\times10^{20} cm^{-2}$ (\cite{2016A&A...594A.116H}). Parameters of this power-law fit are given in Table \ref{tab:AnalysisResult}. The spectra were combined using \href{https://heasarc.gsfc.nasa.gov/ftools/caldb/help/addspec.txt}{addspec} and used in generating SEDs for the two flaring epochs. The X-ray light curve for the source is shown in Figure \ref{fig:LightCurve}. 

\begin{table*}
\caption{The details of \textit{Swift}-XRT analysis.}
\begin{tabular}{|c|c c c c c c|}
    \hline
    Observation ID & Date & Exposure time       & Index & Norm(K) &  F$_{0.2-8 keV}$ & $\chi^{2}_{red.}$\\
     &  DD-MM-YYYY & s & & ph cm$^{-2}$ s$^{-1}$ keV$^{-1}$ & erg cm$^{-2}$ s$^{-1}$ & \\
    \hline
    %00041521004     &07-01-2013  &3969.1  &1.0 &0 &0 &0\\
    00041521005     &12-01-2013  &3628.1  &1.49 &7.91$\times10^{-4}$ &5.84$\times10^{-12}$ &2.146\\
    00049673001     &24-01-2013  &1960.7  &1.39 &7.05$\times10^{-4}$ &5.68$\times10^{-12}$ &1.910\\
    00049673002     &27-01-2013  &2066.0  &1.46 &6.30$\times10^{-4}$ &4.76$\times10^{-12}$ &1.600\\
    00049673003     &02-03-2013  &1213.5  &1.63 &9.21$\times10^{-4}$ &6.06$\times10^{-12}$ &0.584\\
    00049673004     &02-04-2013  &3490.2  &1.58 &8.56$\times10^{-4}$ &5.84$\times10^{-12}$ &1.557\\
    00049673005     &04-04-2013  &3174.2  &1.50 &7.83$\times10^{-4}$ &5.73$\times10^{-12}$ &2.127\\
    00049673007     &06-04-2013  &3490.2  &1.51 &7.95$\times10^{-4}$ &5.76$\times10^{-12}$ &1.771\\
    00049673008     &17-04-2013  &3475.1  &1.46 &7.50$\times10^{-4}$ &5.70$\times10^{-12}$ &1.577\\
    00049673009     &23-04-2013  &2788.1  &1.62 &8.63$\times10^{-4}$ &5.71$\times10^{-12}$ &2.413\\
    00049673010     &29-04-2013  &2319.3  &1.55 &8.72$\times10^{-4}$ &6.13$\times10^{-12}$ &1.617\\
     \hline
\end{tabular}
\end{table*}

\subsection{Optical-UV observation of \textit{Swift}-UVOT}

The optical-UV observations from \textit{Swift}-UVOT \citep{Roming2005} were used in the SEDs modelling and
in estimation of magnetic field in the emission region. 
The observations between MJD 56000-56600 were used in the
present work. The  \textit{Swift}-UVOT obtains the data in six 
filters. Three optical filters in V, B and U bands, and
three UV filters, W1, M2 and W2. For each of the filters, images from different observations were added using \texttt{UVOTIMSUM} tool 
and \texttt{UVOTSOURCE} tool was used to get the magnitude. The Galactic extinction of $E(B-V)=0.1055$ \citep{Schlafly2011} was applied to observed magnitudes which were then converted to
fluxes using Zero point magnitudes \citep{Poole2008}. The fluxes during two flares are mentioned in Table \ref{tab:AnalysisResult}.

\subsection{Optical and infrared observation of SMARTS}
Small $\&$ Moderate Aperture Research Telescope System (SMARTS) is a part of the Cerro Tololo Inter-American Observatory (CTIO) in Chile which observes the sky using visible and IR frequencies. It observes all \textit{Fermi}-LAT monitored blazars that are visible in the Chilean sky in the visible (B, V, R) bands and Near Infrared (J, K) bands. The visible and Near Infrared band fluxes are highly correlated (\cite{2012ApJ...756...13B}).

\begin{figure*}
    \centering
	\includegraphics[width=\textwidth]{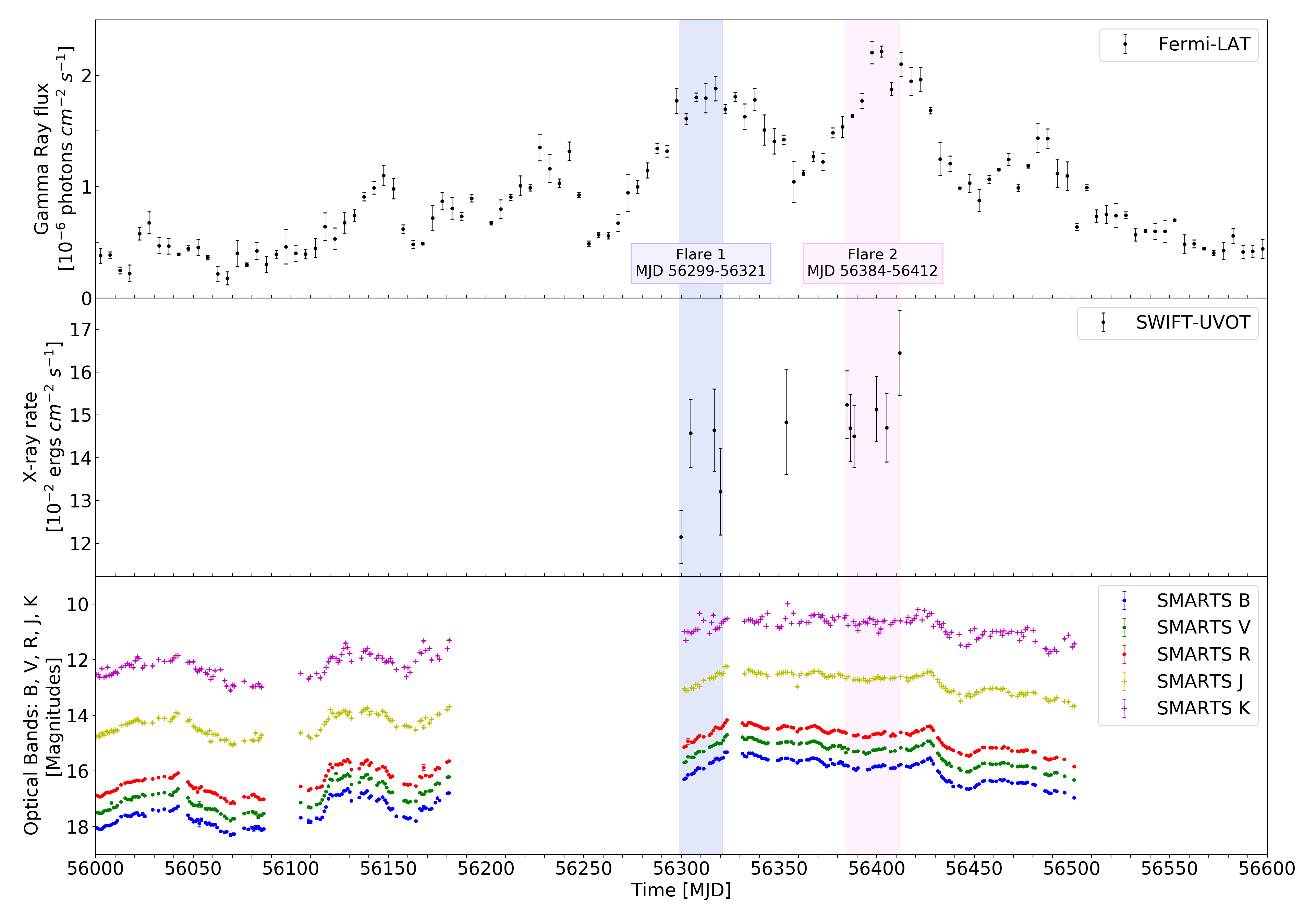}
    \caption[width=\textwidth]{Light curve for the source PKS 1424-418 for all frequencies under consideration.}
    \label{fig:LightCurve}
\end{figure*}

The data for B, V, R, J and K bands was used for the analysis for the time periods MJD 55942-56086, MJD 56104-56181 and MJD 56301-56501. These frequencies are  very close and do not show any major differences in their light curves other than an increase flux at higher frequencies, hence B band was taken as representative of the B, V and R bands group (visible wavelengths) and J band was chosen for representing J and K band group (Near Infrared wavelengths). High correlation between the bands allows for using B band data as a representative of the general nature of the whole visible and IR optical band data. The optical band light curve for the source is shown in Figure \ref{fig:LightCurve} and color index is plotted in Figure \ref{fig:ColorIndex}.

\begin{table*}
    \centering
    \caption{Results from spectral analysis of $\textit{Fermi}$-LAT, $\textit{Swift}$-XRT and $\textit{Swift}$-UVOT data for flare 1 and flare 2}
    \label{tab:AnalysisResult}
    \begin{tabular}{|l| l c c l|}
    \hline
    Instrument & Parameter & Flare 1 & Flare 2 & units \\
    \hline
    \multirow{4}{20mm}{\textit{Fermi}-LAT} & Spectral Index ($\alpha$) & 2.054 & 2.014 & - \\
     & F$_{0.1-300 \ \text{GeV}}$ & $1.77\pm0.04$ & $1.94\pm0.05$ & $10^{-6}\,ph\,cm^{-2}\,s^{-1}$ \\
     & Prefactor ($N_0$) & $3.67\pm0.09$ & $4.17\pm0.10$ & $10^{-10}\,ph\,cm^{-2}\,s^{-1}\,MeV^{-1}$ \\
     & TS & 8251 & 7422 &  - \\
    \hline
    \multirow{4}{20mm}{\textit{Swift}-XRT} & $\Gamma_x$ & $1.41\pm0.07$ & $1.55\pm0.06$ & - \\
     & $K$ & $7.72\pm0.43$ & $8.33\pm0.35$ & $10^{-4}\,ph\,cm^{-2}\,s^{-1}\,keV^{-1}$ \\
     & F$_{0.3-8\,\text{keV}}$ & 6.14 & 5.84 & $10^{-12}\,erg\,cm^{-2}\,s^{-1}$ \\
    \hline
    \multirow{4}{20mm}{\textit{Swift}-UVOT} & v band Flux & $1.467\pm0.064$ & $2.136\pm0.079$ & $ 10^{-11} erg\,cm^{-2}\,s^{-1}$ \\
     & b band Flux & $1.379\pm0.051$ & $1.930\pm0.065$ & $ 10^{-11} erg\,cm^{-2}\,s^{-1}$ \\
     & u band Flux & $1.081\pm0.043$ & $1.608\pm0.058$ & $ 10^{-11} erg\,cm^{-2}\,s^{-1}$ \\
     & w1 band Flux & $0.636\pm0.032$ & $1.027\pm0.046$ & $ 10^{-11} erg\,cm^{-2}\,s^{-1}$ \\
     & m2 band Flux & $0.518\pm0.021$ & $0.816\pm0.033$ & $ 10^{-11} erg\,cm^{-2}\,s^{-1}$ \\
     & w2 band Flux & $0.265\pm0.016$ & $0.370\pm0.019$ & $ 10^{-11} erg\,cm^{-2}\,s^{-1}$ \\
    \hline
\end{tabular}
\end{table*}

\section{Multi-waveband light curve}
In this section we describe the analysis carried out on the multi waveband light curve. Subsection \ref{doubling/halving} quantifies the variability time-scale for the blazar by calculating the doubling/halving time scales. In subsection \ref{correlation} we have studied the cross correlation between the detection of flares in different wavelengths to identify delay in onset of flaring in different bands that might hint at the possible emission mechanism. 

\begin{table*}
    \centering
     \caption[width\textwidth]{Statistically significant doubling time periods for 5-day binned $\gamma$-ray light curve. R(D) stands for Rise(Decay) time. The table is sorted according to increasing doubling/halving time}
    \label{tab:doublingTime}
    \begin{tabular}{|l| c c c c r|}
	    \hline
	    Time bin & Doubling/Halving Time & Rise/Decay & Flux $F(t_1)$& Flux $F(t_2)$ & Statistical Significance \\
	    (MJD) &  (days) &  &  \multicolumn{2}{c}{($10^{-7}$ photons $ cm^{-2} s^{-1}$)} & ($\sigma$) \\
	    \hline
	    56015 - 56020 & 3.61 & (R) & 2.21 & 5.79 & 4.74 \\
	    56065 - 56070 & 4.22 & (R) & 1.77 & 4.02 & 3.85 \\
	    56245 - 56250 & 5.44 & (R) & 9.25 & 4.89 & 19.77 \\
	    56495 - 56500 & 6.43 & (D) & 10.97 & 6.40 & 3.58 \\
	    56055 - 56060 & 6.69 & (D) & 3.64 & 2.17 & 6.94 \\
	    \hline
    \end{tabular}
\end{table*}

\begin{figure*}
    \includegraphics[width=\textwidth]{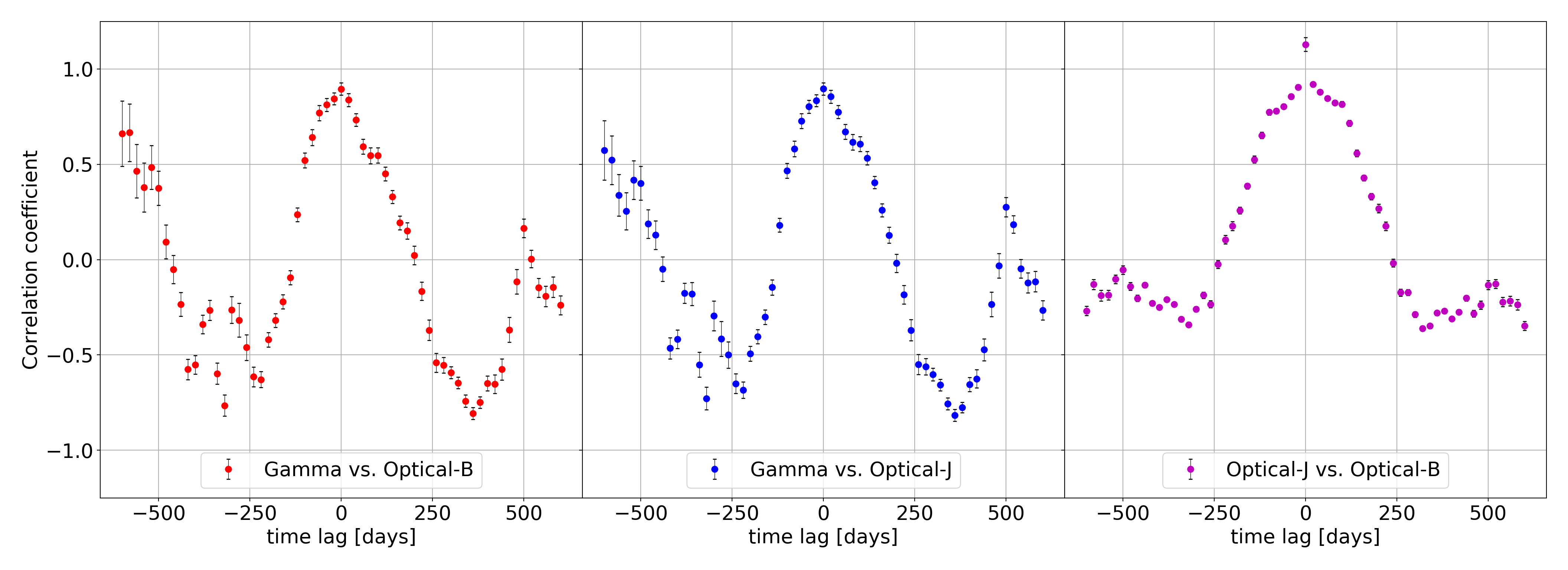}
    \caption[width=\textwidth]{The correlation coefficient as a function of time lag.}
    \label{fig:Correlation1}
\end{figure*}

\subsection{Doubling/halving timescale}
\label{doubling/halving}
Doubling/Halving time periods were calculated for all time bins from MJD 56000 to 56600 for the $\gamma$-ray light curve shown in the Figure \ref{fig:LightCurve}. The computation was carried out using the formula: 
\begin{equation}
    F(t_2) =F(t_1) \times 2^{(t_2-t_1)/T_d}    
\end{equation}
where $F(t_1)$ and $F(t_2)$ are the fluxes at time $t_1$ and $t_2$ respectively, and $T_d$ is the flux doubling/halving time scale. Time periods which had a short $T_d$ with a statistical significance of $\geq3\sigma$ are enumerated in Table \ref{tab:doublingTime} where the significance was calculated using the ratio of the difference in fluxes $F(t_1)$ and $F(t_2)$ and the error in $F(t_1)$. The fastest doubling/halving time scale ($T_{df}$) was found to be 3.61 days with $4.74 \sigma$ confidence which makes $t_{var}$ equal to 2.50 days using $t_{var} = ln(2) \times T_{df}$.

\subsection{Cross-correlation estimations between different wavebands}
\label{correlation}
The correlation between $\gamma$-ray and optical bands was calculated using the Discrete Correlation Function (DCF) as the data points are discrete and the time intervals are not the same for different bands. The DCF allows computation of a correlation coefficient without the use of interpolation for data sampled at different and/or variable rates. The X-ray data was too sparse for a meaningful DCF calculation. The unbinned DCF function (\cite{1988ApJ...333..646E}) can be calculated for two data sets with data point ‘i’ in set 1 and data point ‘j’ in set 2 as:
\begin{equation}
    \text{DCF}(\tau) = \sum_{i,j} \frac{\text{UDCF}_{ij}}{M}
\end{equation} 
\begin{equation}
    \text{UDCF}_{ij}  = \frac{(a_i - \overline{a})\,(b_j - \overline{b})}{\sqrt{(\sigma^2_a-e^2_a)\,(\sigma^2_b-e^2_b)}}
    \label{eq:dcf}
\end{equation}
where $\tau$ is the DCF bin size, $\Delta t_{ij}=(t_j-t_i)$ is the lag for the pair ($a_i,b_i$) and M is the total number of pairs for which \\$(\tau - \Delta\tau/2) \leq \Delta  t_{ij}  \leq (\tau + \Delta\tau/2)$. $\overline{a}$ and $\overline{b}$ are the averages of $a_i$ and $b_i$ respectively. $\sigma$ and $e$ are the standard deviation and measurement error associated with each set. The error in DCF can be calculated as:
\begin{equation}
    \sigma_{\text{DCF}(\tau)}=\frac{1}{M-1} \sqrt{\sum_{i,j}(\text{UDCF}_{ij} - \text{DCF}(\tau))^2} 
\end{equation}

It was found that the optical and $\gamma$-ray flux is highly correlated with zero lag based on the maxima of the correlation coefficient plots (Figure \ref{fig:Correlation1}). The correlation coefficient peaked at zero time lag for B band vs $\gamma$-ray, J band vs $\gamma$-ray and B band vs J band. The unavailability of X-ray data for most of the time period under consideration resulted in insufficient amount of data points for X-ray vs $\gamma$-ray correlation analysis. The correlation coefficient is plotted with time lags in Figure \ref{fig:Correlation1}.

\subsection{Flux-Index correlation}
\begin{figure}
    \centering
	\includegraphics[width=0.6\columnwidth]{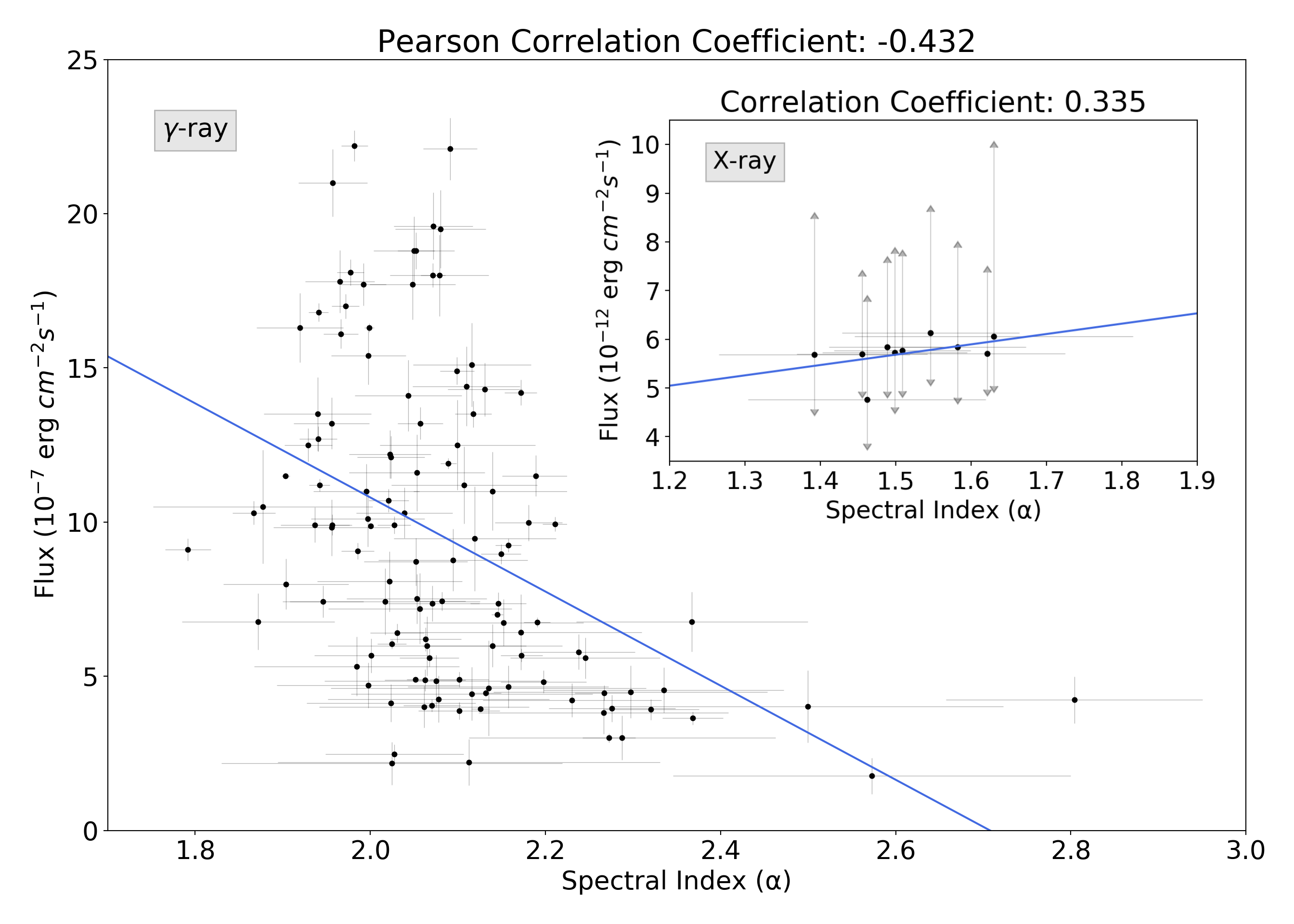}
    \caption[width=\textwidth]{The Flux-Index plot for \textit{Fermi}-LAT data. A correlation was observed between high flux and low power law spectral index ($\alpha$). The slope is negative and Pearson Correlation Coefficient is -0.432. The scatter in the data points is too high for a meaningful linear regression, the blue line given is a linear fit for reference. The inset plot is Flux vs Index for \textit{Swift}-XRT data. The slope is positive and Pearson Correlation Coefficient is 0.335, which is opposite of the trend observed in the $\gamma$-ray Flux-Index plot. The blue line is a linear fit for reference}
    \label{fig:Index_hardening}
\end{figure}

We computed the Flux Index and Color Index correlations for the multi waveband data to analyse the variability. The $\gamma$-ray flux vs index plot is shown in Figure \ref{fig:Index_hardening}. A correlation is observed between high flux and low spectral index ($\alpha$). This type of index hardening previously observed in other blazars during their high flux states (\cite{2007ApJ...657..706N}). The X-ray flux vs Index plot (an inset to Figure \ref{fig:Index_hardening}) showed an inverse trend compared to the $\gamma$-ray plot, the flux was higher at larger values of spectral index.

\begin{figure*}
	\includegraphics[width=\textwidth]{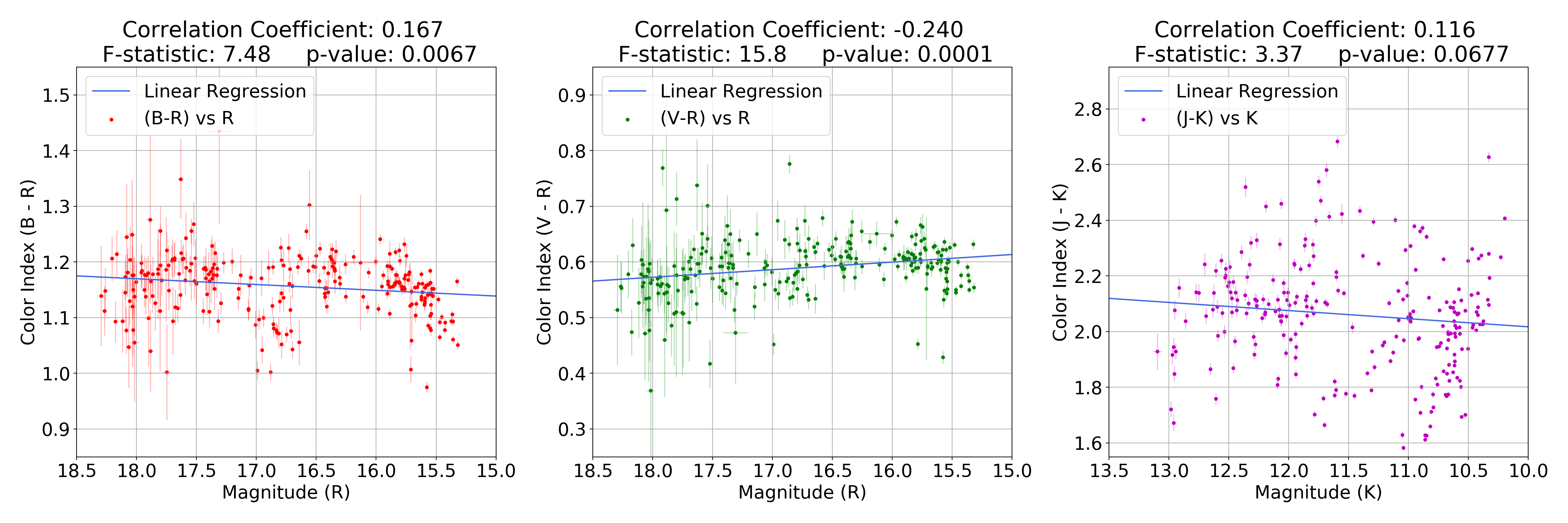}   
    \caption[width=\textwidth]{The Color Index plots for SMARTS optical band data.The Pearson correlation coefficient value is very small, indicating no correlation between the quantities in x and y axes. Null hypothesis for the F-statistic test was a constant fit and the alpha-level was 0.05. This suggests that the null hypothesis can be safely rejected for (B-R) vs R and (V-R) vs R plots since their p-values are smaller than alpha and F-stat values are much greater than 1 but the same cannot be said for the (J-K) vs K plot where p-value is larger than alpha level.}
    \label{fig:ColorIndex}
\end{figure*}

Color Index vs Magnitude plots allow for the determination of relative strengths of the thermal emission from the accretion disk and the non-thermal emission in the jet. The color index plots were analysed further for statistical significance of linear fits. F-statistic test was carried out, with a constant fit being the null hypothesis and a linear fit being the alternate hypothesis. The F-statistic values and p-values for (B-R) vs R, (V-R) vs R and (J-K) vs K color index plots strongly suggest that the null hypothesis is invalid and there exists a correlation between Color index and magnitude of a reference band. The scatter and error bars are large which reduces significance of the numerical value of the slope of the plots and hence only the nature of the correlation bears meaning. The (B-R) vs R and (J-K) vs K color index plots in Figure \ref{fig:ColorIndex} show redder-when-brighter trend while the (V-R) vs R plot shows a bluer-when-brighter trend. 

\section{Spectral energy distribution and its modelling}
\begin{figure}
    \centering
	\includegraphics[width=0.6\columnwidth]{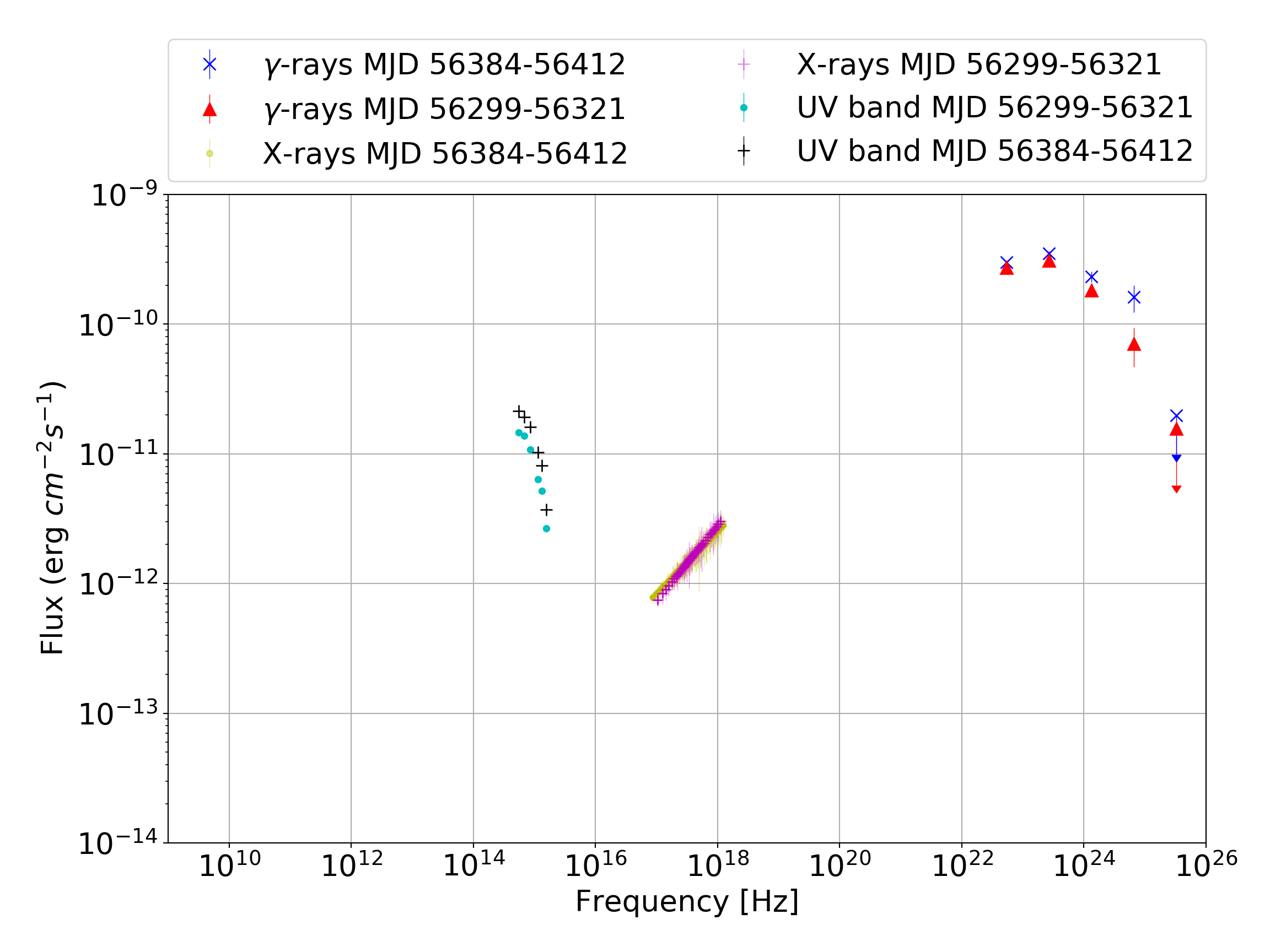}
    \caption[width=\textwidth]{The spectral energy distribution of the source, combined plot for UV, X-ray and $\gamma$-ray bands for the time periods indicated in the legend}
    \label{fig:SED_data}
\end{figure}

The simultaneous multi waveband SEDs were generated for two flaring periods 1 \& 2 to understand the broadband emissions from PKS 1424-418. 
The model fitting was done using a publicly available code JetSet\footnote{https://jetset.readthedocs.io/en/latest/} \citep{2011ApJ...739...66T, 2009A&A...501..879T, 2006A&A...448..861M}. The model assumes the spherically symmetric blob having radius $(R')$ and an entangled magnetic field $(B)$. The blob makes an angle of $\theta$ with the observer and moves with the bulk Lorentz factor of $\Gamma$. This makes the emitted flux from the blob to be affected by the beaming factor, $\delta = 1/\Gamma(1- \beta \cos \theta)$. The blob is filled with a relativistic population of electrons following an empirical lepton distribution relation. A broken power law model of the form 

\begin{equation}
N(\gamma) = 
    \begin{cases} 
        K_1\times\gamma^{\alpha_1} & \quad for \quad \gamma_{min} \leq \gamma \leq \gamma_{break}\\
        K_2\times\gamma^{\alpha_2} & \quad for \quad \gamma_{break} < \gamma \leq \gamma_{max}
    \end{cases} 
\end{equation}

was used for the lepton distribution where $\gamma$ = $E/mc^2$ is the Lorentz factor, $\gamma_{min}$, $\gamma_{break}$ and $\gamma_{max}$ are model parameters, $\alpha_1$ $\&$ $\alpha_2$ are the spectral indices below and above break Energy $E_b$ respectively and $K_2 = K_1 (\gamma_{break})^{\alpha_2-\alpha_1}$ where $K_1 \& K_2$ are normalisation constants (for detailed discussion on the broken power law model, see \cite{2001A&A...367..809K}).

These relativistic electrons interact with magnetic fields in the emission region and produce synchrotron photons in the frequency region of the first hump of the SED. The emission in the frequency region of the second hump is reproduced either by SSC or EC model. In case of SSC model the seed photons for inverse Compton process are the synchrotron photons produced by the same population of relativistic electrons. For the EC model, the seed photons are the following external photon fields:
\begin{itemize}
    \item The direct emission in Optical region from the accretion disk 
    \item The reprocessed emission in Optical-UV region from the BLR \citep{Donea2003}
    \item The reprocessed emission in IR region from dusty torus
\end{itemize}
In this work we have modeled spectra of PKS1424-418 with seed photons from the dusty torus, explained in detail in the next section.

\begin{figure*}
	\includegraphics[width=\textwidth]{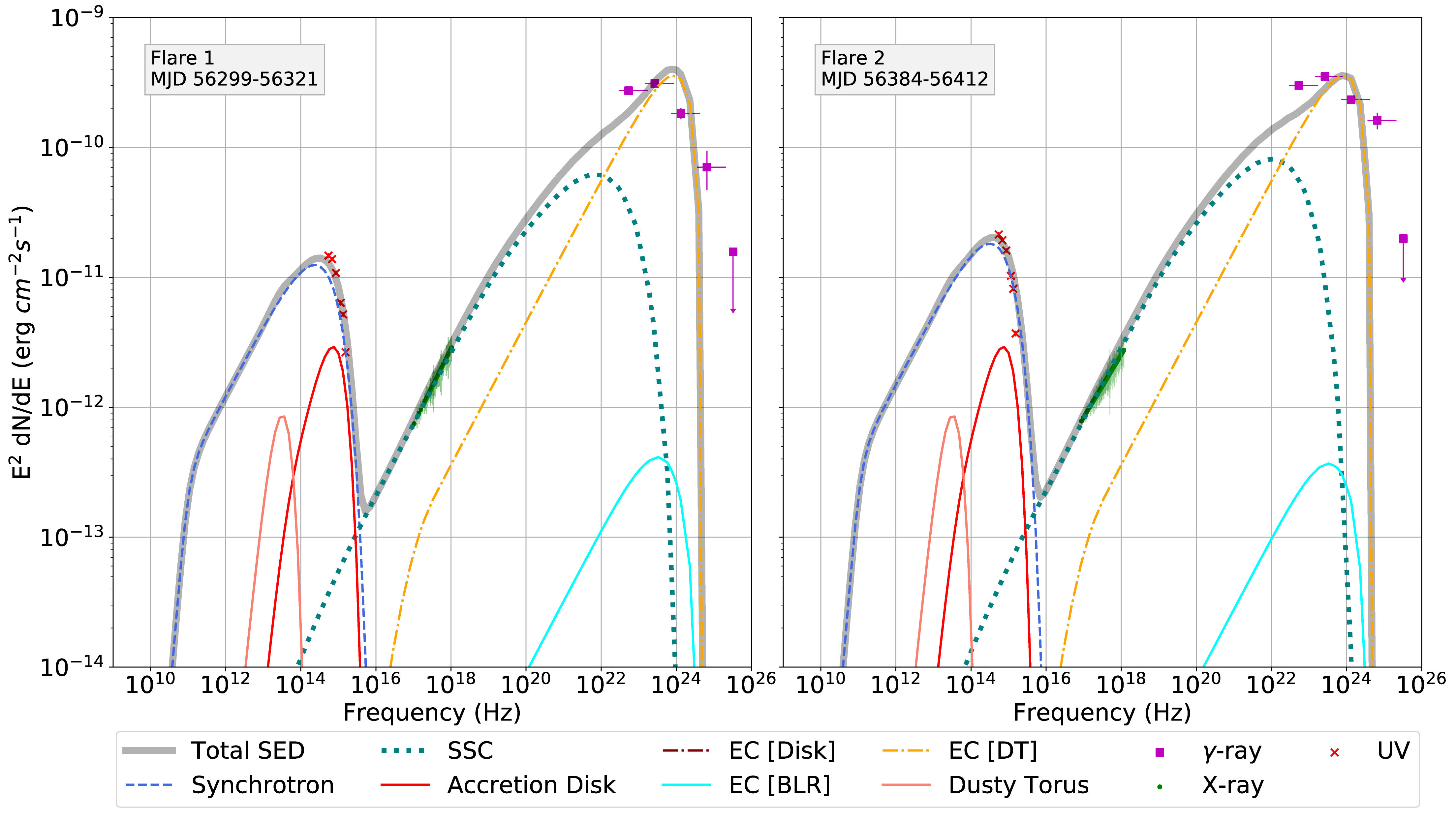}
    \caption[width=\textwidth]{Spectral energy distribution of observational data and fitted model for the two flaring periods}
    \label{fig:SED_fit_both_flares}
\end{figure*}

The disk emission is being modelled as multi-temperature black body radiation given by 
\begin{equation}
T^4(R) = \frac{3R_s L_{disk}}{16 \epsilon \pi \sigma_{SB} R^{3}} \left[ 1 - \left(\frac{3R_s}{R}\right)^{1/2} \right]
\end{equation}

where $L_{disk}$ is the disk luminosity, $R_s$ is the Schwarzschild radius and $\sigma_{SB}$ is the Stefan-Boltzmann constant. The value of $L_{disk}$ used in modelling of average state of PKS 1424-418 in the population study of blazars by \cite{2017ApJ...851...33P} and \cite{Celotti2008} was (1.0-5.0) $\times$ 10$^{47}$ erg s$^{-1}$. With these values of $L_{disk}$ and assuming scaling relation \citep{10.1111/j.1365-2966.2009.15898.x}, the distance of BLR from central black hole is estimated as, $(1.0 - 2.24) \times 10^{18}$ cm. In our model, we have used the value of $L_{disk}$ and black hole mass as $9.0 \times 10^{46} \text{erg} \, s^{-1}$ and $4.5 \times 10^{9} \, M_\odot$ \citep{2004ApJ...602..103F}, respectively. While modelling both the flaring states, the parameters external to the jet were kept fixed. The parameters of electron distribution and Magnetic field ($B$) of emission region were varied, keeping $R'$, $\Gamma$ and $\theta$ fixed. The parameters used to model the flaring state SEDs are given in Table \ref{tab:SEDJetSeT} and modelled SEDs are shown in Figure \ref{fig:SED_fit_both_flares}.

\section{Result and discussion}
In this section, the results of temporal and cross-correlation analysis, constrained on Doppler factor, estimation of size of emission region and broadband modelling of SEDs are discussed.

The significant correlation at near-zero time lag was found during MJD 56000 to 56600 between, $\gamma$-ray emission and the optical-IR emission. Also, the optical and IR emissions were found to have significant correlation without any observable time lag. This suggests that the $\gamma$-ray, optical and IR emission regions are co-spatial within uncertainty of the average time gap of data points. For PKS 1424-418, \cite{2014A&A...569A..40B} also found a good correlation between optical and GeV emission for the three flares that occurred during 2008-2011. A similar tightly correlated variation at near-zero lag between optical and IR bands was also found in long-term study of FSRQ 3C 454.3 \citep{Sarkar2019}.

\subsection{The constraint on Doppler factor}

Using the detection of the highest energy photons from the source, the value of minimum Doppler factor can be estimated. This estimation assumes the optical depth, $\tau_{\gamma\gamma} (E_h)$, of the highest energy photon, $E_h$, to $\gamma\gamma$ interaction is 1. It is given by 
\begin{equation}
\delta_{min} = \Bigg[ \frac{\sigma_{t} d_{l}^{2} (1+z)^{2} f_{\epsilon} E_{h}}{4t_{var}m_{e}c^{4}} \Bigg]^{1/6},    
\end{equation}
where, $\sigma_{t}$ is Thomson cross-section, $d_{l}$ is the luminosity 
distance which is $11.55$ Gpc for this source, $f_{\epsilon}$ is the X-ray flux in 0.3-8.0 keV energy range, $E_{1}$ is the energy of the highest energy photon, and $t_{var}$ is the observed variability timescale.
The fastest variability time scale ($t_{var}$) was found  to be $2.50$ days in present work. The highest energy photon observed during the first flare was $38.33$ GeV, while it was $32.31$ GeV during the second flare. The probabilities of these high energy photons being associated with PKS 1424-418 were $>$99.98$\%$ and $>$99.99$\%$, respectively. Using the X-ray fluxes as given in Table \ref{tab:AnalysisResult}, the values of $\delta_{min}$ were estimated as 12.5 and 12.0, during the first and second flare, respectively. \cite{2014RAA....14.1135F} estimated this limit as $13.39$ and $17.87$ using average energy of $\gamma$-ray photon as 4.91 GeV, the authors used a variability timescale of one day and 6 hrs, respectively, for simplicity.

\subsection{The size of emission region}
The $\gamma$-ray light curve (Figure \ref{fig:LightCurve}) was scanned for estimation of the doubling/halving time during the period from MJD 56000-56600, which was used to calculate the size of emission region.
The size of the emission region, $R'$ was estimated using the following equation:
\begin{equation}
    R' = c \, \delta_{min} \, t_{var}/(1+z)
\end{equation}

$R'$ was estimated to be $(3.11-4.64) \times 10^{16}$ cm, using the $\delta_{min}$ range of $12.0-17.87$, as mentioned in the previous section. In SED modelling $R'$ of $5.0 \times 10^{16}$ cm was used.

\begin{table*}
\centering
\caption{The input parameters used to reproduce the observed broadband SEDs of the two flaring states}
\label{tab:SEDJetSeT}
\begin{tabular}{|l| l c c c|}
\hline
 Sr. No. & Model parameters & Unit   & Flare 1 & Flare 2   \\
\hline
1. & $R'$               & 10$^{16}$ cm  &5.0     &5.0       \\
2. & $R_{BLR}$          & 10$^{18}$ cm           &2.0-2.3    &2.0-2.3     \\
3. & $\theta$           & degree        &3.05      &3.05      \\   
4. & $\Gamma$           & -             &20.0      &20.0      \\  
5. & $\delta$           & -             &18.76      &18.76      \\
6. & $d$                & 10$^{18}$ cm  &9.0      &9.0\\
7. & B                  & G             &0.22   &0.28      \\	
8. & $\gamma_{min}$     & -             &1.9      &1.5      \\
9. & $\gamma_{break}$   & -             &8$\times10^{3}$      &9$\times10^{3}$      \\
10. & $\gamma_{max}$     & -             &$10^{4}$      &$10^{4}$      \\
11.& $N$              & cm$^{-3}$     &7000      &7500      \\         
12.& ${\alpha_1}$       & -             &1.9      &1.9      \\
13.& ${\alpha_2}$       & -             &4.1      &4.1      \\
14.& U$'_e$             & erg cm$^{-3}$ &0.132    &0.117           \\
15.& U$'_B$             & erg cm$^{-3}$ &0.0019     &0.003        \\	
16.& U$'_e$/U$'_B$      & -             &69.47     &39    \\
17.& M$_{BH}$           & M$_\odot$     &4.5$\times10^{9}$         &4.5$\times10^{9}$   \\
18.& $L_{d}$            & 10$^{46}$ erg s$^{-1}$ &9.0    &9.0     \\
19.& $T_{DT}$           & K             & 1000  & 1000  \\
20.& $R_{DT}$   & 10$^{19}$ cm  &2.0      &2.0\\
21.& $\tau_{DT}$     & -     & 0.2       & 0.2\\
\hline
\end{tabular}

\begin{tablenotes}
\item Note: [1] The size of emission region in co-moving frame
            [2] BLR radius
            [3] Viewing angle
            [4] Bulk Lorentz factor
            [5] Doppler beaming factor
            [6] The distance of emission region from the black hole, calculated using $R'$ and $\Gamma$
            [7] Magnetic field
            [8-10] Minimum, break and maximum Lorentz factor of injected electron spectrum
            [11] Particle density
            [12] Low energy particle spectral index
            [13] High energy particle spectral index
            [14] Electron energy density in co-moving frame
            [15] Magnetic field energy density in co-moving frame
            [16] Equipartition value
            [17] Mass of central SMBH
            [18] Accretion disk luminosity
            [19] Temperature of the Dusty Torus
            [20] Radius of the Dusty Torus
            [21] Fraction of disc luminosity reprocessed in IR emission
\end{tablenotes}	
\end{table*}

\subsection{Broadband emission during flaring states}

Two major flaring periods occurred with the time gap of around three months in early 2013, were modelled with a single zone leptonic scenario. The peak flux during both flares was $\sim 2 \times 10^{-6}$ ph cm$^{-2}$ s$^{-1}$. Figure \ref{fig:SED_fit_both_flares} shows the modelled SEDs of both the flaring periods. The modelled parameters are mentioned in Table \ref{tab:SEDJetSeT}. The model parameters during both the flares were similar within an order of magnitude. This suggests the similar nature of particle acceleration mechanism, resulting in low energy and high energy particle spectral indices of $1.9$ and $4.1$ respectively for both the flares. During these flares the optical-UV emission is dominated by synchrotron emission from the jet rather than the thermal emission from disk as seen in well-know FSRQ 3C 273 \citep{doi:10.1142/S2010194512004886}. 

The X-ray emission was reproduced with the SSC process. Highest energy photons detected during these flares are above 30 GeV. This implies that the $\gamma$-ray emission happens beyond broad line region, as this region is highly opaque. Thus the $\gamma$-ray emission was modelled using the EC mechanism with photons from the dusty torus, as done by \cite{Tavecchio2013}. The flares exhibited by the source during 2008-2011 were modelled by \cite{2014A&A...569A..40B} with a single zone EC model. The authors reported that similar model parameters could explain the SEDs during all four flares, which is in line with our conclusion based on the two 2012-2013 flares. 

The reported spectral indices for Flare B1 and B2 in \cite{2014A&A...569A..40B} are similar to the spectral indices in this work. They have modelled the SED using contributions from EC BLR and EC Dusty Torus while in our case the dominant external photon contribution is from the Dusty Torus based on the fact that we have observed very high energy photons and hence the $\gamma$-ray emission region is likely outside the BLR since it is opaque to such photons. Our work predicts a different equipartition value, \cite{2014A&A...569A..40B} have come to the conclusion that the Magnetic field carries most of the power with electrons carrying a fraction of the total power of the jet for the 2008-11 flares while we have arrived at the opposite result for the 2012-13 flares, suggesting electrons carry much of the energy. This is because of the fact that the magnetic field predicted by our fitting is almost 10 times smaller ($\sim 0.2$ gauss) compared to the $2.5$ gauss magnetic field in \cite{2014A&A...569A..40B}. Their Bulk Lorentz factor ($\Gamma$) and Doppler factors ($\delta$) are almost twice the values obtained in our SED fitting.

\section{Summary}
PKS 1424-418 was found to be in flaring state during the time periods MJD 56299-56321 and MJD 56384-56412 based on the $\gamma$-ray light curve, further analysis was carried out for these time periods. The $\gamma$-ray, optical and IR bands were found to be highly correlated with zero time lag suggesting the same emission region. The shortest doubling/halving time period was found to be $3.6$ days from the light curve analysis (Table \ref{tab:doublingTime}) which is consistent with the general timescale of the variability of FSRQs. The modelling of the source suggests physical parameters of the jet as enumerated in Table  \ref{tab:SEDJetSeT}, the nature of the two flares was found to be similar based on the comparison of parameters obtained in the modelling. The observation of the highest energy photon of $38.33$ GeV and $32.31$ GeV during these flares constrained the Doppler factor to be $>12.0$ and suggests that the $\gamma$-ray emission occurs outside the BLR region.

\section*{Acknowledgements}
We thank the referee for useful and constructive comments which helped in improving the manuscript. 

This work has used data obtained by Fermi \&  Swift missions and SMARTS telescopes. This research made use of Enrico, a community-developed Python package to simplify \textit{Fermi}-LAT analysis \cite{2013ICRC...33.2784S}. D Bose acknowledges the support of Ramanujan fellowship.

\section*{Data Availability}
For this work we have used data from \textit{Fermi}-LAT, \textit{Swift}-XRT,  \textit{Swift}-UVOT and \textit{SMARTS} telescopes. All the data used is available in the public domain. Details are given in section 2. 

%\bibliography{PKS1424}
%\bibliographystyle{mnras}  

\end{document}